\documentstyle[aps,prl,epsfig,floats]{revtex} 
\begin{document}
\draft 
\wideabs{
\title  {Critical currents in Josephson junctions, with unconventional pairing symmetry: $d_{x^2-y^2}+is$ versus $d_{x^2-y^2}+id_{xy}$}
\author{ N. Stefanakis, N. Flytzanis}
\address{Department of Physics , University of Crete,
	P.O. Box 2208, GR-71003, Heraklion, Crete, Greece}

\date{\today}
 
\maketitle

\begin{abstract}
Phenomenological Ginzburg-Landau theory is used to
calculate the possible spontaneous vortex states that may exist
at corner junctions of $d_{x^2-y^2}+ix$-wave, (where $x=s$ or $x=d_{xy}$)
and $s$-wave superconductors. We study 
the magnetic flux and the critical current modulation 
with the junction orientation angle $\theta$, the magnitude 
of the order parameter, and the magnetic field $H$.
It is seen that the critical current $I_c$
versus the magnetic flux $\Phi$ relation is symmetric / asymmetric
for $x=d_{xy}/s$ when the orientation is exactly
such that the lobes of
the dominant $d_{x^2-y^2}$-wave order parameter points towards the two
junctions, which are at right angles for the corner junction.
The conclusion is that a measurement of the $I_c(\Phi)$ relation
may distinguish which symmetry ($d_{x^2-y^2}+is$ or
$d_{x^2-y^2}+id_{xy}$) the order parameter has.
\end{abstract}

}

\section{Introduction}
One of the main questions in the research
activity on high-$T_c$ superconductors nowadays is the identification
of the order parameter symmetry and its underlying mechanism\cite{scalapino,vanh}. 
The most possible scenario is that the 
bulk pairing state has a $d_{x^2-y^2}$-wave character. 
Theoretical calculations, suggest
that an imaginary $s$-wave component which breaks the time reversal 
symmetry is induced in some cases, wherever the $d_{x^2-y^2}$-wave 
order parameter varies spatially such as near a vortex, 
or near the surface
\cite{shiba}.
Also the observation of fractional vortices on a triangular grain boundary in 
YBa$_2$Cu$_3$O$_7$ by Kirtley $et$ $al.$ \cite{kirtley2}, 
may indicate a possible violation 
of the time-reversal symmetry near grain boundary.
Theoretical explanation of this experiment is given by Bailey et. al. in Ref. \cite{bailey}
where they study a triangular grain boundary in $d$-wave 
superconductors. They conclude that under the assumption of 
$d$-wave symmetry, the flux at the edges of this triangle 
can take the values $\pm \Phi_0/2$, which does not agree with the 
experiment.  
However under the assumption of $d_{x^2-y^2}+is$-wave symmetry  
an intrinsic phase shift $\phi_c(x)$ exists in each 
triangle edge. In turn the phase $\phi(x)$ 
must change in order to connect the different values of
$\phi_c$ in each segment. This arrangement leads to 
fractional vortices or antivortices at each three corners, 
in agreement with the experiment.

Another pairing state which breaks the time reversal symmetry is 
the $d_{x^2-y^2}+id_{xy}$-wave.
Patches of complex $d_{xy}$ components are induced around 
magnetic impurities at low temperatures in a $d_{x^2-y^2}$-wave superconductor
forming a phase coherent state as a result of tunneling between 
different patches \cite{balatsky1}. 
Violation of parity and time reversal symmetry occurs in this state.
Also on the high field region, $H\le H_{c2}$ the $d_{x^2-y^2}$-wave 
state can be perturbed by the external filed, producing a 
$d_{x^2-y^2}+{\it i}d_{xy}$ state in the bulk \cite{balatsky2}.

The observation of the splitting of the zero energy  
peak in the conductance spectra at low temperatures indicates 
that a secondary component is 
induced which violates locally the time reversal symmetry 
\cite{covington}. Theoretical 
explanation based on surface-induced Andreev states, 
has been proposed \cite{fogelstrom}.
Recently the field dependence of this splitting has been 
observed in the tunneling spectra of YBCO \cite{aprili,krupke}.
This observation
is consistent with a $d_{x^2-y^2}+is$ surface order parameter
as well as a $d_{x^2-y^2}+id_{xy}$ bulk order parameter. 
Another question which can be asked is to what extend, the 
observation of a symmetric magnetic interference pattern in 
the corner junction experiments \cite{vanh} is an identification 
of $d_{x^2-y^2}$-wave symmetry, or could also imply a 
$d_{x^2-y^2}+id_{xy}$ pairing state also?
In this work we propose a phase sensitive experiment 
based on the Josephson effect, which may be used to 
distinguish which symmetry ($d_{x^2-y^2}+is$ or $d_{x^2-y^2}+id_{xy}$) 
the order parameter has near
the surface.
We study the static properties of a frustrated
junction which is made of two one-dimensional junctions,
of $d_{x^2-y^2}+ix$-wave, (where $x=d_{xy}$
or $x=s$) and $s$-wave superconductors.
By introducing an extra relative phase in one part of this 
junction, the above junction can be mapped into the corner junctions 
experiments\cite{vanh,yanoff}.
We examine the spontaneous flux and the critical current modulation 
of the vortex states with the 
junction orientation angle $\theta$, the magnitude 
of the secondary component $n_s$, and the magnetic field $H$.
In each case we derive simple arguments which are useful to 
discriminate between the time reversal symmetry broken states.
For example, when the orientation is exactly
such that the lobes of
the dominant $d_{x^2-y^2}$-wave order parameter points towards
the junction interface the magnetic interference pattern
is symmetric (asymmetric) when the secondary order parameter
is $x=d_{xy}(s)$. 
This is verified for small junctions
as well as in the long junction limit, and can be used to 
distinguish between broken time reversal symmetry states.  

The rest of the paper is organized as follows. In Sec. II we
discuss the Josephson effect between a superconductor with 
broken time reversal symmetry and an $s$-wave superconductor.
In Sec. III the geometry of the corner junction is 
discussed. In Sec. IV 
we present the results for the magnetic flux of the spontaneous 
vortex states in corner junctions with some intrinsic magnetic flux. 
In Sec. V the parameters which can modulate the spontaneous 
flux and the critical currents are considered. 
In Sec. VI a connection with the experiment is made.
Finally, a summary and discussion are presented in the last section.

\section{Josephson effect between two superconductors with  mixed wave symmetry} 
\begin{figure}
  \centerline{\psfig{figure=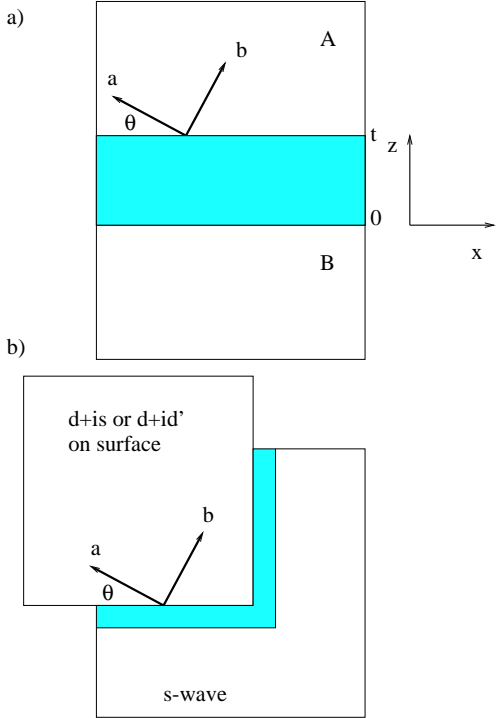,width=6.5cm,angle=0}}
\caption{
(a) A single Josephson junction between superconductors $A$ and 
$B$ with a two component order parameter. The angle between the 
crystalline $a$ axis of $A$ and the junction interface is 
$\theta$. (b) The geometry of the corner junction between 
a mixed symmetry superconductor, and an $s$-wave superconductor.}
  \label{fig1.fig}
\end{figure}

We consider the junction shown in Fig. \ref{fig1.fig}(a), where two superconductors 
($A$ in the region $z>t$ and $B$ in the region $z<0$), 
are separated by 
an intermediate layer. We assume that each superconductor has a  
two component order parameter. The order parameter for each 
component $k (k=1,2)$ in the superconductors, can be written as
\begin{equation} 
  n_k = \left\{ 
    \begin{array}{ll}
      \widetilde{n}_k^Ae^{{\it i} \phi_k^A}  
, & z > t \\
      \widetilde{n}_k^Be^{{\it i} \phi_k^B}  
, & z < 0
    \end{array}.~~~\label{ni}
\right.
\end{equation}
Here $\phi_k^{A(B)}$ is the phase of the order 
parameter $n_k$ in superconductor $A(B)$. Then phenomenological Ginzburg-Landau theory is used to
calculate the supercurrent density given by
\cite{zhu}
\begin{equation}
J=\sum_{k,l=1}^2 J_{ckl}\sin(\phi_k^B-\phi_l^A),~~~\label{j}
\end{equation}
where 
\begin{equation}
\begin{array}{ll}
J_{c11} = & (2 e \hbar / m_a^{*} d) \widetilde{n}_1^A \widetilde{n}_1^B\\
J_{c21} = & (2 e \hbar / m_{\nu}^{*} d) \widetilde{n}_1^A \widetilde{n}_2^B\\
J_{c12} = & (2 e \hbar / m_{\nu}^{*} d) \widetilde{n}_2^A \widetilde{n}_1^B\\
J_{c22} = & (2 e \hbar / m_b^{*} d) \widetilde{n}_2^A \widetilde{n}_2^B 
\end{array},~~~\label{jkl}
\end{equation}
$m_a^{*}, m_{\nu}^{*}, m_b^{*}$ are the effective masses that enter into the
Ginzburg-Landau equations. In the following these masses are taken equal to 
an effective mass $m^{*}$.

We restrict to the case where $B$ is $s$-wave. In this case 
$\widetilde{n}_1^B=0$, and 
$\widetilde{n}_2^B$=constant.
We define 
$\phi=\phi_2^B-\phi_1^A$, as the relative phase difference 
between the two superconductors. We consider the case where the intrinsic 
phase difference within superconductor $A$ is $\phi_2^A-\phi_1^A=
\pi/2$. Then the order parameter in $A$ is complex and breaks the 
time reversal symmetry. The supercurrent density can be written as:

\begin{equation}
J(\phi)=\widetilde J_c \sin(\phi+ \phi_c),~~~\label{jphidis}
\end{equation}
with 
\begin{equation}
\widetilde 
J_c=\sqrt{J_1^2+J_2^2},~~~\label{jcdis}
\end{equation}

\begin{equation}
  \phi_c = \left\{ 
    \begin{array}{ll}
      \tan^{-1}\frac{J_2}{J_1}, & J_1 > 0 \\ 
      \pi + \tan^{-1}\frac{J_2}{J_1}, & J_1 < 0  
    \end{array},~~~\label{phic}
\right.
\end{equation}
where $J_1=J_{c21}$, $J_2=-J_{c22}$. The Josephson critical current 
density $\widetilde J_c$ is scaled in units of 
$J_{c0}=\frac{e \hbar}{m^{*} d}$.
Two special cases are the following:

i) For $d_{x^2-y^2}+{\it i}s$ wave case 
the magnitude of the
$d_{x^2-y^2}$-wave component in (\ref{ni}) is $\widetilde{n}_1^{A}=
n_{10}\cos(2\theta)$ 
, where $\theta$ is the angle of the crystalline $a$-axis
of superconductor $A$ with the junction interface.
The magnitude of the secondary order parameter in superconductor 
$A$ is $\widetilde{n}_2^{A}=
n_{20}=0.1n_{10}$.

ii) For $d_{x^2-y^2}+{\it i}d_{xy}$ wave case, 
the magnitude of the $d_{x^2-y^2}$-wave component 
in (\ref{ni}) is given by $\widetilde{n}_1^{A}=
n_{10}\cos(2\theta)$, while the $d_{xy}$ wave component is 
$\widetilde{n}_2^{A}=n_{20}\sin(2\theta)$, where $n_{20}=0.1n_{10}$. 
This order parameter
can occur in the following way: The order parameter magnitude for the 
$d$-wave state $\Delta_0(\theta)=\Delta_0\cos(2\theta)$ 
is an equal admixture of pairs with orbital moment 
$L_z=\pm2$, and can be written as $\Delta_0(\theta)=
(\Delta_0/2)[\exp(2{\it i}\theta)+\exp(-2{\it i}\theta)]$.
In the presence of perturbation such as (ferromagnetically) ordered impurity
spins $S_z$ the coefficients of $L_z=\pm2$ components will shift 
linearly in $S_z$ with opposite signs. The final state will be
$\Delta_0(\theta)\rightarrow
\Delta_0(\theta)+{\it i}S_z\Delta_1(\theta)$, where 
$\Delta_1(\theta)=\sin(2\theta)$. The strength of the secondary 
component is proportional to the perturbation $S_z$.

\section{The corner junction geometry} 

We consider the corner junction shown in Fig. \ref{fig1.fig}(b), between 
a superconductor with broken time reversal symmetry at the surface 
and an s-wave superconductor.
If the angle of $a$-axis with the interface in the $x$-direction 
is $\theta$, then the corresponding angle in the $z$-direction 
will be $\pi/2-\theta$. We map the two segments each of length $L/2$
where $L=10\lambda_J$ of this junction 
into a one-dimensional axis. In this case the two dimensional junction can 
be considered as being made of two one dimensional junctions described 
in Sec. II connected in parallel. Their characteristic phases 
$\phi_{c1}$ and $\phi_{c2}$ depend upon the angle $\theta$.
We call this junction frustrated since the two segments have 
different characteristic phases $\phi_{c1}, \phi_{c2}$.
The fabrication details of 
corner junctions or superconducting quantum interference device (SQUID), 
between sample faces at different angles
can be found in Ref. \cite{vanh,yanoff}.

The superconducting phase difference $\phi$ across the junction is 
then the solution of the sine-Gordon (s-G) equation
\begin{equation} 
  \frac{d^2 { \phi}(x)}{dx^2} = \widetilde J_c \sin[{
\phi(x)+\phi_c(x)}] - I^{ov},~~~\label{eq01} 
\end{equation}
with the boundary conditions
\begin{equation} 
  \frac{d \phi}{dx}|_{x=0,L} = H
.~~~\label{eqbc} 
\end{equation}
The length $x$ is scaled in units of the 
the Josephson penetration depth given by 
\[
\lambda_J=\sqrt{\frac{\hbar c^2}{8\pi e d J_{c0}}},
\] 
where $d$ is the sum of the $s$-wave, and mixed wave $\lambda_{ab}$  
penetration depths plus the thickness of the 
insulator layer.
The relative phase 
$\phi_c(x)$ is $\phi_{c1} (\phi_{c2})$ in the left (right) part of the
junction. 
The external magnetic field $H$, scaled in units of $H_c=\frac{\hbar c}{2 e d \lambda_J}$
is applied in the $y$ direction, which 
is considered small compared to $\lambda_J$.
The bias current per unit length $I^{ov}$ in the overlap geometry 
is scaled in units of $\frac{c}{4\pi}H_c$, 
and is uniformly distributed 
along the entire $x$ axis of the junction. 

We can classify the different solutions obtained from 
Eq. (\ref{eq01}) with their magnetic flux content 
\begin{equation}
  \Phi = \frac{1}{2 \pi}  (\phi_R-\phi_L) ,~~~\label{phi} 
\end{equation}
where $\phi_{R(L)}$ is the value of $\phi$ at the right(left) edge of 
the junction, in units of the flux quantum
$\Phi_0= \frac{h c}{2 e}$.

\section{Spontaneous vortex states} 

\begin{figure}
  \centerline{\psfig{figure=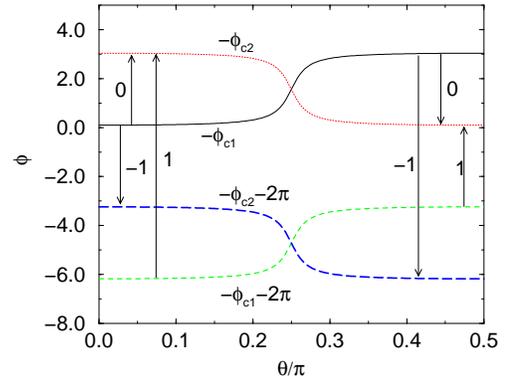,width=6.5cm,angle=0}}
\caption{
The stable solutions $\phi = - \phi_{c1}+2n_1\pi$
($\phi = - \phi_{c2}+2n_2 \pi$), for $n_i=0,-1$, $i=1,2$, that exist in the 
left(right) junction,
of $d_{x^2-y^2}+is$-wave and $s$-wave superconductors, when considered 
uncoupled, at zero current, versus the orientation angle $\theta$. 
Each junction has length $L/2$, where $L=10\lambda_J$, and 
$\phi_{c1}, (\phi_{c2})$
is the extra phase difference in the left (right) junction due to the 
different orientations.
The arrows denote the variation of the phase $\phi$ in order to 
connect these stable solutions in the frustrated junction geometry.
We present three possible solutions i.e. 
$n=0,-1,1$, and 
down(up) arrow denotes 
negative(positive) magnetic flux. 
}
\label{fig2.fig}
\end{figure}

Firstly let us consider the case where the two one-dimensional 
junctions of $d_{x^2-y^2}+ix$-wave where $x=s$ or $x=d_{xy}$,
and $s$-wave supeconductors, 
each of length $L/2$, described in Sec. II are uncoupled. 
Then for $0<x<L/2$ the stable solutions for the s-G equation are 
$\phi(x)=-\phi_{c1}+2n_1 \pi$, where $n_1=0,\pm 1,\pm 2,...$, 
while for 
$L/2<x<L$ the stable solutions for the s-G equation are
$\phi(x)=-\phi_{c2}+2n_2 \pi$, where $n_2=0,\pm1,\pm2,...$, 
where  $\phi_{c1}$, $\phi_{c2}$, are the relative phases in 
each part of the junction due to different orientations.
These solutions are plotted in Fig. \ref{fig2.fig}, for $n_i=0, -1$, $i=1,2$
as a function of the orientation angle $\theta$.
When the frustrated junction is formed, and we consider the 
above junctions in parallel, the phase $\phi$ is forced to 
change around $x=L/2$, to connect these stable solutions. 
This variation of the phase $\phi$, along the junction 
describes the Josephson vortices. 
The flux content of these states (in units of $\Phi_0$) is \cite{sigrist}
\begin{equation}
\Phi=[\phi(L)-\phi(0)]/2\pi = 
(-\phi_{c2}+\phi_{c1}+2n\pi)/2\pi,~~~\label{con}
\end{equation}
where the $n$-value ($n=n_1-n_2=0,\pm 1,\pm 2,...$)
distinguishes between solutions with 
different flux content. We will 
concentrate to solutions called modes with the minimum flux content i.e., 
$n=0, 1, -1$. 
Their magnetic flux in terms of $\phi_{c1}, \phi_{c2}$ 
is shown in table \ref{n=01_1}.
Generally the flux content is fractional i.e. is neither integer 
nor half-integer, as a consequence of the broken time reversal symmetry
of the problem.

\begin{table}
\caption{
The magnetic flux ($\Phi$) in terms of $\phi_{c1}$, $\phi_{c2}$
for the spontaneous solutions that exist in the corner junction 
geometry between a superconductor with time reversal broken symmetry
and an $s$-wave superconductor ($\phi_{c1}, \phi_{c2}$
is the extra phase difference in the two edges of the corner junction due to the
different orientations, of the $a$-axis of the dominant $d_{x^2-y^2}$-wave 
superconductor). We present only the minimum 
flux states $n=0,-1,1$.
}
\begin{tabular}{cc}
Vortex state $n$ & Magnetic flux ($\Phi$)\\ \tableline
$0$ & $(-\phi_{c2}+\phi_{c1})/2\pi$\\
$1$ & $(-\phi_{c2}+\phi_{c1}+2\pi)/2\pi$\\
$-1$ & $(-\phi_{c2}+\phi_{c1}-2\pi)/2\pi$\\ 
\end{tabular}
~~~\label{n=01_1}
\end{table}

In the actual numerical simulations, the stable solutions of 
the sine-Gordon equation in the left(right) part of the 
junction are taken as the initial conditions for the iteration 
procedure. For example for the $n=0$ 
solution the phase $\phi(x)$ is taken $\phi(x)=-\phi_{c1}$ 
$(-\phi_{c2})$ in the left (right) part of the junction, as an 
initial condition and then is iterated until convergence. 
Besides if we take as initial condition 
, $\phi(x)=-\phi_{c1}$, in the left side, and 
$\phi(x)=-2\pi-\phi_{c2}$ in the right side, 
the final state of the system, after the 
iteration procedure, is the solution which we call $n=-1$, 
with negative magnetic flux, and not exactly opposite 
to $n=0$. We comment here that the solutions after the 
iteration procedure have smooth variation as a function 
of the position, as opposed to the step function variation 
of the initial conditions.

For the $0-0$ junction $\phi_{c1}=\phi_{c2}=0$, and the flux becomes 
$\Phi=n$, so we say that the flux is quantized in integer units 
of $\Phi_0$.
In this case, there exist solutions with flux 
$\Phi=...,-1,0,1,...$ \cite{caputo}. These solutions, 
when $n \neq 0$ are stabilized 
by the application of an external magnetic field. 
In the case of a junction with some spontaneous flux, 
at least for the modes with lower flux content, 
the external field is not necessary 
since the spontaneous magnetization state is stable.

In the case of $0-\pi$ junction, where the intrinsic phase in the 
right (left)
part of the junction is $\phi_{c2}=-\pi$ ($\phi_{c1}=0$), the stable
solutions of the s-G equation are $\phi(x)=2n\pi$ for the left part, while 
$\phi(x)=\pi(2n+1)$ for the right part of the junction. In this case
a $0-\pi$ junction is formed. The corresponding flux
becomes 
$\Phi=(n+1/2)\pi$, and the particular values of $n=0$, $n=-1$ 
give the half vortex and antivortex solutions, with opposite fluxon 
content, $\Phi=0.5$ and $\Phi=-0.5$ respectively. 

\section{Magnetic flux and critical current modulation} 

In the following we will describe three parameters which 
can alter the spontaneous flux and the critical currents 
of the vortex states described 
in the previous section, in a corner junction between a superconductor 
with time reversal broken symmetry and an $s$-wave superconductor. 
These include the orientation angle $\theta$, 
the magnitude of the secondary order parameter $n_s$,
and the magnetic field $H$. In each parameter separately 
we will point out the differences between the 
$d_{x^2-y^2}+is$-wave, and $d_{x^2-y^2}+id_{xy}$-wave.
\subsection{Junction orientation}
\begin{figure}
  \centerline{\psfig{figure=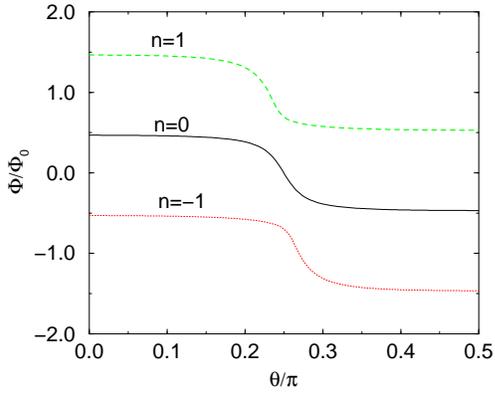,width=6.5cm,angle=0}}
\caption{
The magnetic flux $\Phi$ as a function of the
angle $\theta$, for the various vortex states,
$n=0,-1,1$, that exist spontaneously 
in a corner junction between a
$d_{x^2-y^2}+is$-wave and an $s$-wave superconductor, with length
$L=10\lambda_J$. The flux for $\theta=0$
is fractional.
}
\label{fig3.fig}
\end{figure}

\begin{figure}
  \centerline{\psfig{figure=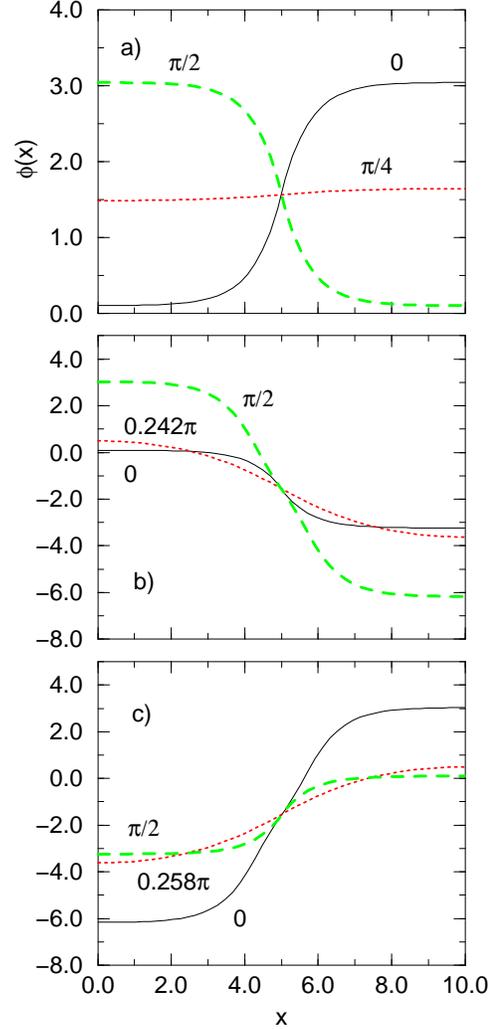,width=6.5cm,angle=0}}
\caption{The phase distribution of the vortex solutions a) $n=0$, 
at $\theta= 0$, $\pi /4$, $\pi /2$;
b) $n=-1$, at $\theta= 0$, $0.242 \pi$, where the instability sets in,
and $\pi /2$;
c) $n=1$, at $\theta=0$, $0.258 \pi$, at the point where the 
instability occurs, and $\pi /2$, 
for a corner junction of $d_{x^2-y^2}+is$-wave and $s$-wave 
superconductors, with length 
$L=10\lambda_J$, and zero overlap external current $I^{ov}=0$.}
\label{fig4.fig}
\end{figure}

\begin{figure}
  \centerline{\psfig{figure=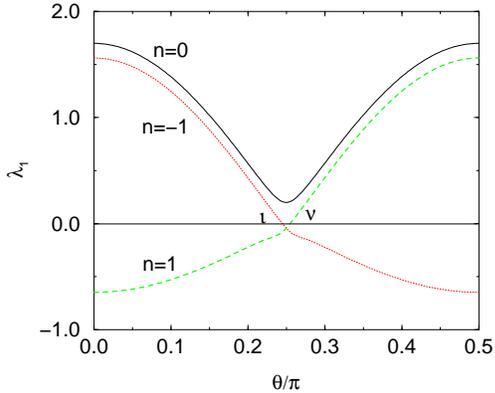,width=6.5cm,angle=0}}
\caption{
The lowest eigenvalue $\lambda_1$ of the linearized eigenvalue 
problem
as a function of angle $\theta$, for the $n=0,-1,1$
solutions. In the range where $\theta$ is close to zero, the eigenvalues 
for both 
$n=0$, and $-1$ are positive and correspond to stable solutions.
}
\label{fig5.fig}
\end{figure}

For the $d_{x^2-y^2}+{\it i}s$-wave case, we consider first the situation where $\theta$ 
is varied from $0$ to $\pi/2$. 
In Fig. \ref{fig3.fig} we plot the spontaneous magnetic flux versus the angle
($\theta$) for the different modes $n=0,-1,1$
in the corner junction geometry.  
As we can see the magnetic flux changes with orientation.
For angle $\theta$ close to $0$ or $\pi/2$ the spontaneous modes 
existing at $H=0$ are separated by an integer value of the 
magnetic flux. This is also the case in the pure $s$-wave 
superconductor  junction 
problem. The difference is that the modes are found displaced 
to fractional values of magnetic flux, contrary to the 
$s$-wave case where the magnetic flux takes on integer 
values at $H=0$.  
In particular the vortex solution in the $n=0$ mode (solid line) contains
less that half a fluxon for $\theta=0$, and as we increase the 
angle $\theta$ towards $\pi /4$ it continuously reduces its flux, 
i.e. it becomes flat exactly at $\theta=\pi /4$ and then it 
reverses its sign and becomes an antivortex with exactly opposite
flux content at $\theta=\pi /2$ from that at $\theta=0$.
In addition we have plotted in Fig. \ref{fig4.fig}a 
the phase distributions for the mode $n=0$ in different 
orientations $\theta=0, \pi /4, \pi /2$. The transition form the vortex 
to the antivortex mode as the orientation changes is clearly 
seen in this figure.
Note that the solutions in this mode remain stable for all 
junction orientations. This is seen in Fig. \ref{fig5.fig} where we plot 
the lowest eigenvalue ($\lambda_1$) of the linearized eigenvalue
problem as a function of the angle $\theta$ \cite{caputo}. 
We see that $\lambda_1>0$, denoting stability for all values 
of the angle $\theta$ in this mode.

Let as now examine the solution in the $n=-1$ mode, (dotted line 
in Fig. \ref{fig3.fig}). We see that at 
$\theta=0$ it has negative flux, which in absolute value is 
more than $\Phi_0 /2$  and as we increase the angle $\theta$ 
it decreases its flux to a full antifluxon when the 
orientation is slightly greater than $\pi /4$ and than to flux 
greater than $\Phi_0$ when $\theta$ reaches $\pi /2$. 
As seen in Fig. \ref{fig5.fig} this solution becomes unstable at a point 
to the left of $\theta= \pi /4$ (point $\iota$) due to the abrupt change of 
flux at this angle. More strictly the instability sets in 
due to the competition between the slope of the phase at 
the edges of the junction and at the junction center as the
angle $\theta$ approaches the value $\pi /4$. At this point 
the slope competition makes 
the antivortex unstable. 
This is seen in Fig. \ref{fig4.fig}b) (dotted line)
where the phase distribution for the 
$n=-1$ mode solution is plotted 
at the point where the instability 
starts i.e. $\theta=0.242 \pi$.

Finally the solution in the $n=1$ mode contains more than one 
fluxon at $\theta=0$ and is clearly unstable. It becomes stable 
at an angle slightly on the right of $\theta=\pi /4$, (point $\nu$ in 
Fig. \ref{fig5.fig}) 
where 
the flux varies more smoothly, see
$\theta=0.258 \pi$ in Fig. \ref{fig4.fig}c.
At $\theta=\pi /2$ it contains more than $\Phi_0 /2$ in flux. 
We expect a time reversal broken symmetry state like 
$d_{x^2-y^2}+is$ to be characterized by either 
the solution in the fractional vortex or antivortex mode, 
because due to the different character of these solutions 
a change from one variant to the other would demand the application 
of an external current or magnetic field and in this sense it 
would cost additional energy. So since these states are stable in external 
perturbations, once the system is prepared in one of 
these it will remain to that state. 

\begin{figure}
  \centerline{\psfig{figure=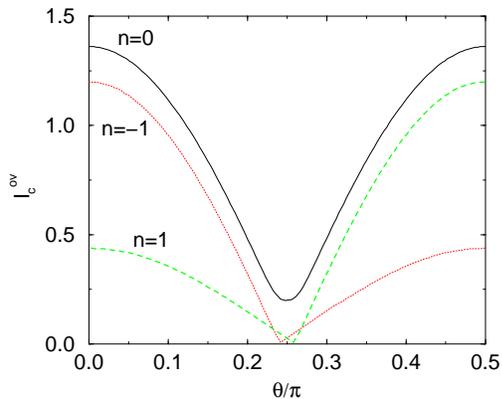,width=6.5cm,angle=0}}
\caption{
Overlap critical current $I_c^{ov}$ per unit length versus the angle $\theta$
for a corner junction of $d_{x^2-y^2}+is$-wave and $s$-wave superconductors,
with length
$L=10\lambda_J$, 
for the vortex solutions $n=0,-1,1$
that exist spontaneously in the junction.
}
\label{fig6.fig}
\end{figure}

\begin{figure}
  \centerline{\psfig{figure=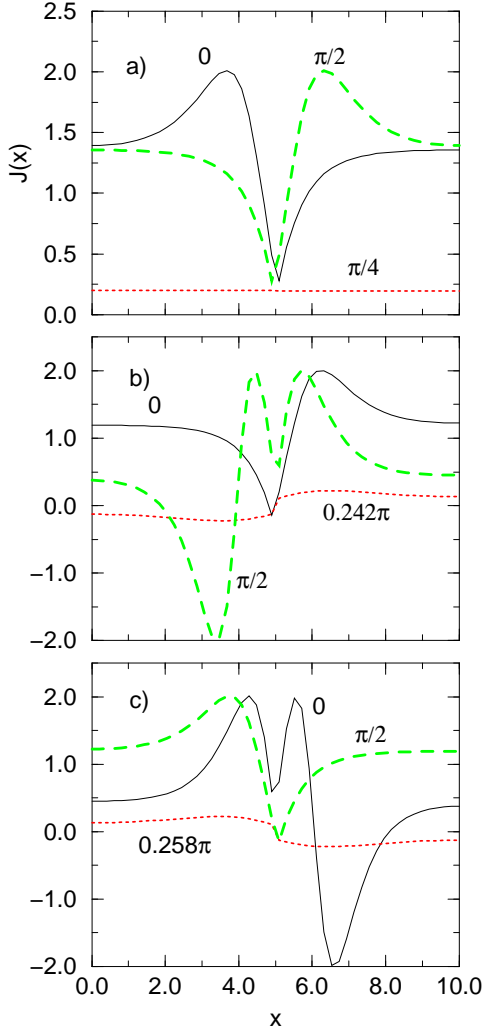,width=6.5cm,angle=0}}
\caption{The current density distribution $J(x)$ 
of the vortex solutions a) $n=0$, 
at $\theta= 0$, $\pi /4$, $\pi /2$, 
b) $n=-1$, at $\theta= 0$, $0.242 \pi$, where the instability sets in, 
and $\pi /2$, 
c) $n=1$, at $\theta=0$, $0.258 \pi$, at the point where the 
instability occurs, and $\pi /2$, 
for a corner junction of $d_{x^2-y^2}+is$-wave and $s$-wave 
superconductors, with length 
$L=10\lambda_J$, and maximum external overlap current $I_c^{ov}$.}
\label{fig7.fig}
\end{figure}

In general we see that for each value of $\theta$ there exist in 
the junction a pair of stable solutions which when 
applying an external bias current will lead 
to observable critical currents.
In Fig. \ref{fig6.fig} we plot the overlap critical current per unit 
length $I_c^{ov}$ as a function 
of $\theta$, at $H=0$, for the $n=0,-1,1$-mode solutions, in 
the $d_{x^2-y^2}+is$-wave case. 
In the overlap geometry the current is distributed in the entire $x$-axis.
In the calculations we have taken into account that the Josephson critical 
current density $\widetilde J_c$ has a characteristic variation with 
the orientation. 
We find that 
for a given orientation 
it is possible for the junction current density to vary in the 
way that several modes with different critical currents 
can exist.
In Fig. \ref{fig7.fig} we plot 
the current density when the total current is maximum, for different 
modes, and orientations,  
which will give us information about the actual shapes 
of the vortices.
Let us consider the situation where the 
junction contains a solution in the mode $n=0$, at $\theta=0$,  
when the net current is maximum.
The spatial variation of $\phi$ is described by a fractional vortex 
which is displaced around the value $\phi= \pi$, from the 
corresponding distribution at zero current which is around $\pi /2$
(see Fig. \ref{fig4.fig}a). 
The current density distribution as seen in Fig. \ref{fig7.fig}a (solid line)
at the maximum current is flat 
above unit with a small variation around the junction center giving 
rise to the large value on the net current, seen in Fig. \ref{fig6.fig}. 
Also at $\theta= \pi/4$ the flat phase distribution corresponding 
to the $n=0$ solution at zero current is displaced towards the 
value $\phi= \pi$ when applying an external current.
The corresponding current distribution seen in Fig. \ref{fig7.fig}a, (dotted line)  
is straight line and the net current is small 
for this orientation.
For the $n=-1$ solution at the point where the instability sets in 
i.e. $\theta=0.242 \pi$, 
the current density distribution is symmetric around zero as 
seen in Fig. \ref{fig7.fig}b (dotted line) and carries 
zero net current at this point. Thus the instability occurs just   
before the angle where a full antifluxon enters the junction. A 
slightly different situation occurs in the magnetic interference 
pattern of a pure $s$-wave superconductor junction \cite{owen} 
where, the net current is zero at the 
magnetic field where a full fluxon or antiluxon enters the junction, 
in the no flux $0$-mode.
At the point $\theta= \pi/2$, of the $n=-1$-mode the junction 
contains more than
one fluxon causing the characteristic oscillations in the current
density around the junction center as seen in Fig. \ref{fig7.fig}b
(dashed line).
This reduces the critical current 
for this orientation.

\begin{figure}
  \centerline{\psfig{figure=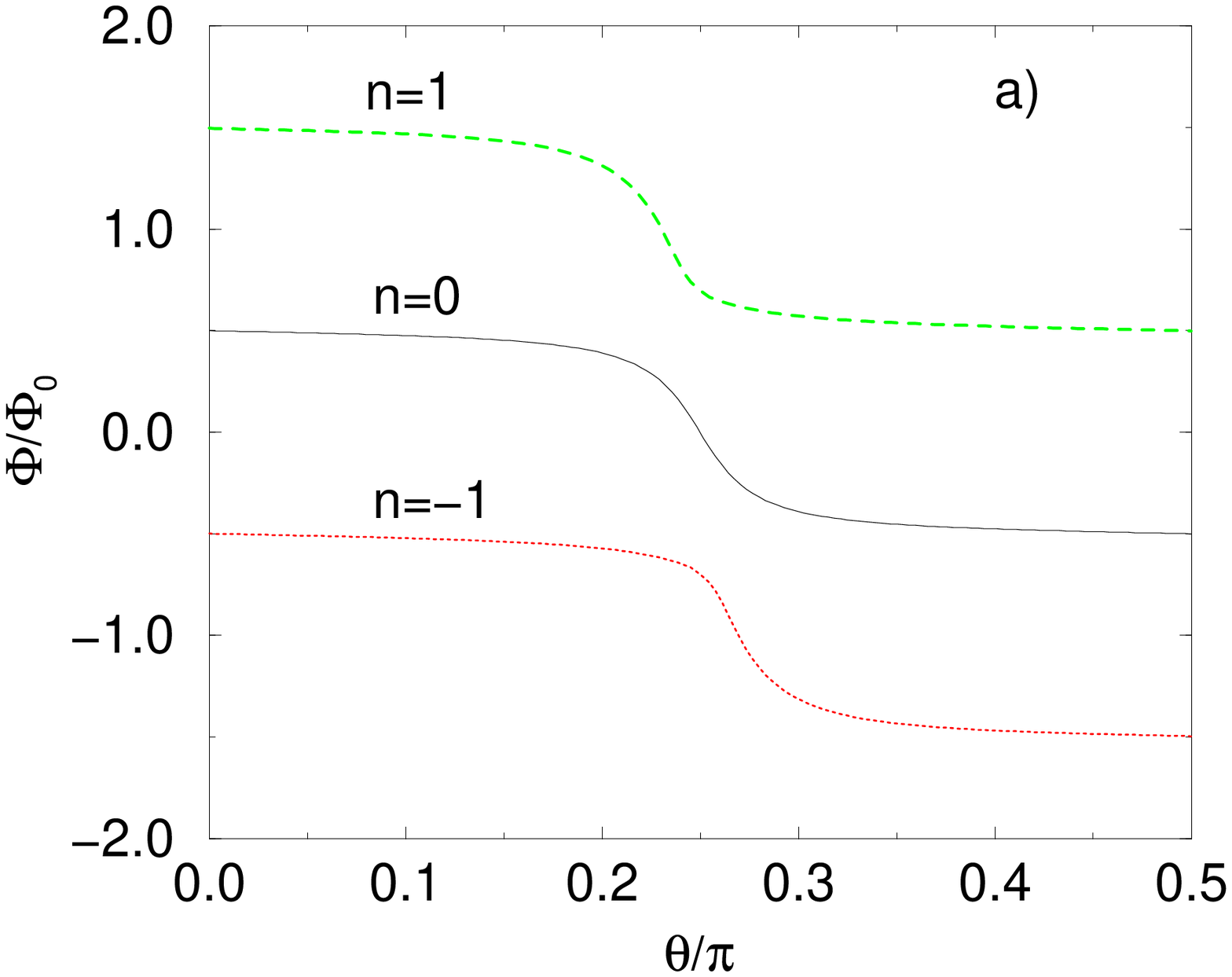,width=6.5cm,angle=0}}
  \centerline{\psfig{figure=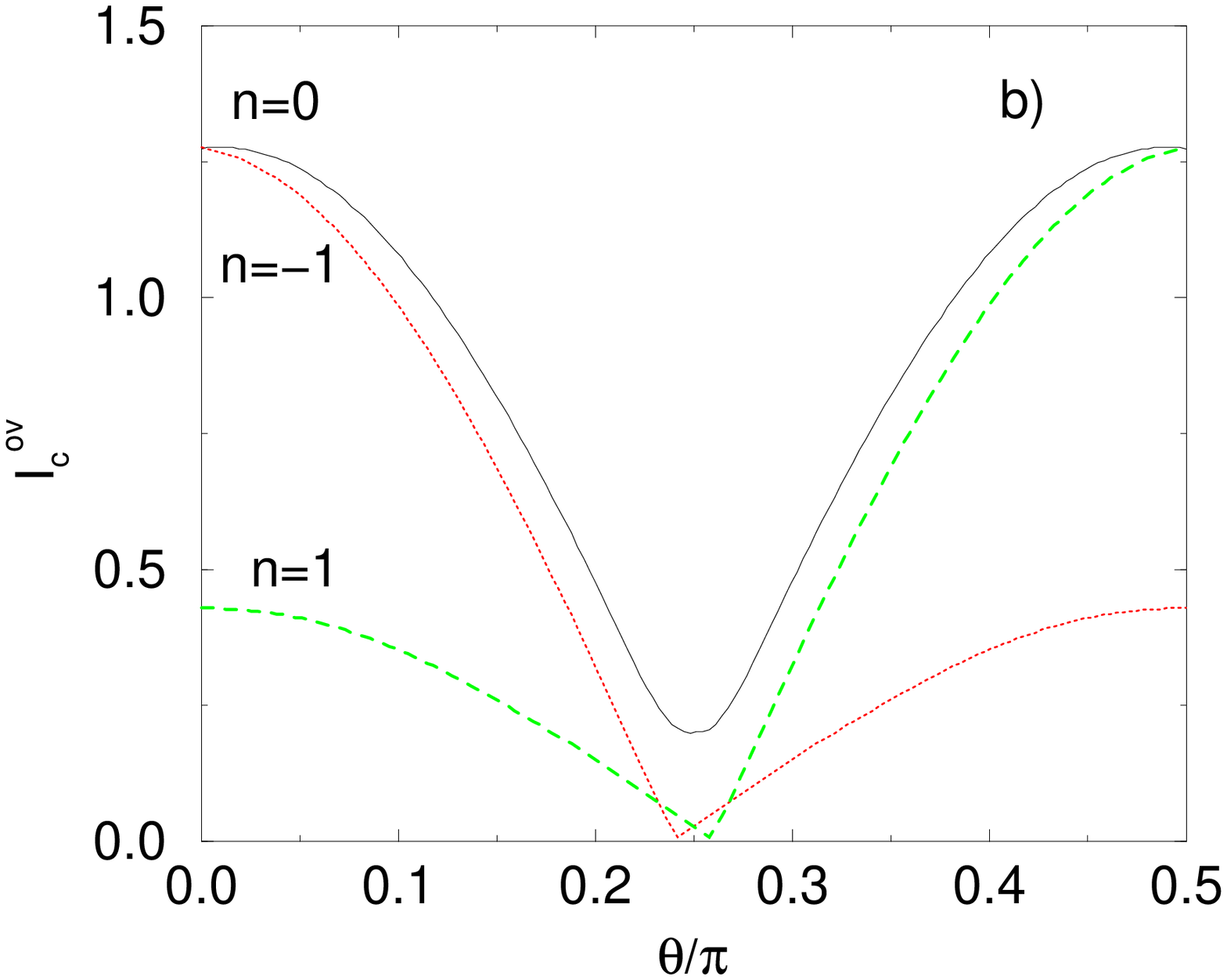,width=6.5cm,angle=0}}
\caption{
a) The spontaneous magnetic flux $\Phi$ as a function of the
angle $\theta$, for the various vortex states,
$n=0,-1,1$, for a corner junction between a
$d_{x^2-y^2}+id_{xy}$-wave and an $s$-wave superconductor, with length
$L=10\lambda_J$. The flux for $\theta=0$
is integer multiply of $\Phi_0 /2$.
b) The corresponding critical current $I_c^{ov}$ per unit length.
}
\label{fig8.fig}
\end{figure}

For the $d_{x^2-y^2}+{\it i}d_{xy}$ pairing symmetry state, we plot in 
Fig. \ref{fig8.fig}a) the flux content for the $n=0,-1,1$, versus the 
angle $\theta$.
Note the half integer or multiplies value of $\Phi$ at $\theta$ 
close to $0$ or $\pi/2$.
For this grain orientation
the magnetic flux is only sensitive to the real part of the 
order parameter, which has a sign change but does not break 
time-reversal symmetry.
In the $d_{x^2-y^2}+is$-wave state the order 
parameter is complex for all junction orientations and 
breaks the time-reversal symmetry. Close to $0$ or $\pi/2$ 
the flux is fractional.
The flux quantization at $\theta=0$ can be used to 
discriminate between 
these states. 

In Fig. \ref{fig8.fig}b) we plot the critical current per unit length 
evolution 
with the grain angle $\theta$ in the 
$d_{x^2-y^2}+{\it i}d_{xy}$-wave state. Close to $\theta=0$ we see that
the $I_c^{ov}$ for the $n=0,-1$ solutions, coincide.
This 
happens also at $\theta=\pi/2$ for the $n=0,1$ solutions.
In these orientations the order parameter becomes pure real 
and does not break the time-reversal symmetry. As a result
the critical current at these angles is the same as in a junction with 
pure $d$-wave symmetry.
At $\theta=\pi /4$ the order parameter is pure imaginary 
and has the same magnitude 
for both pairing states.
As a consequence 
for $\theta=\pi/4$, the critical currents for both junctions 
coincide.
Also the unstable part of the $n=1$ branch, in the $I_c$ vs $\theta$
is almost the same for the two symmetry states, due to the 
small difference in the flux, compared with the 
large flux content of the solutions in this region.

\subsection{Magnitude of the secondary order parameter}
\begin{figure}
  \centerline{\psfig{figure=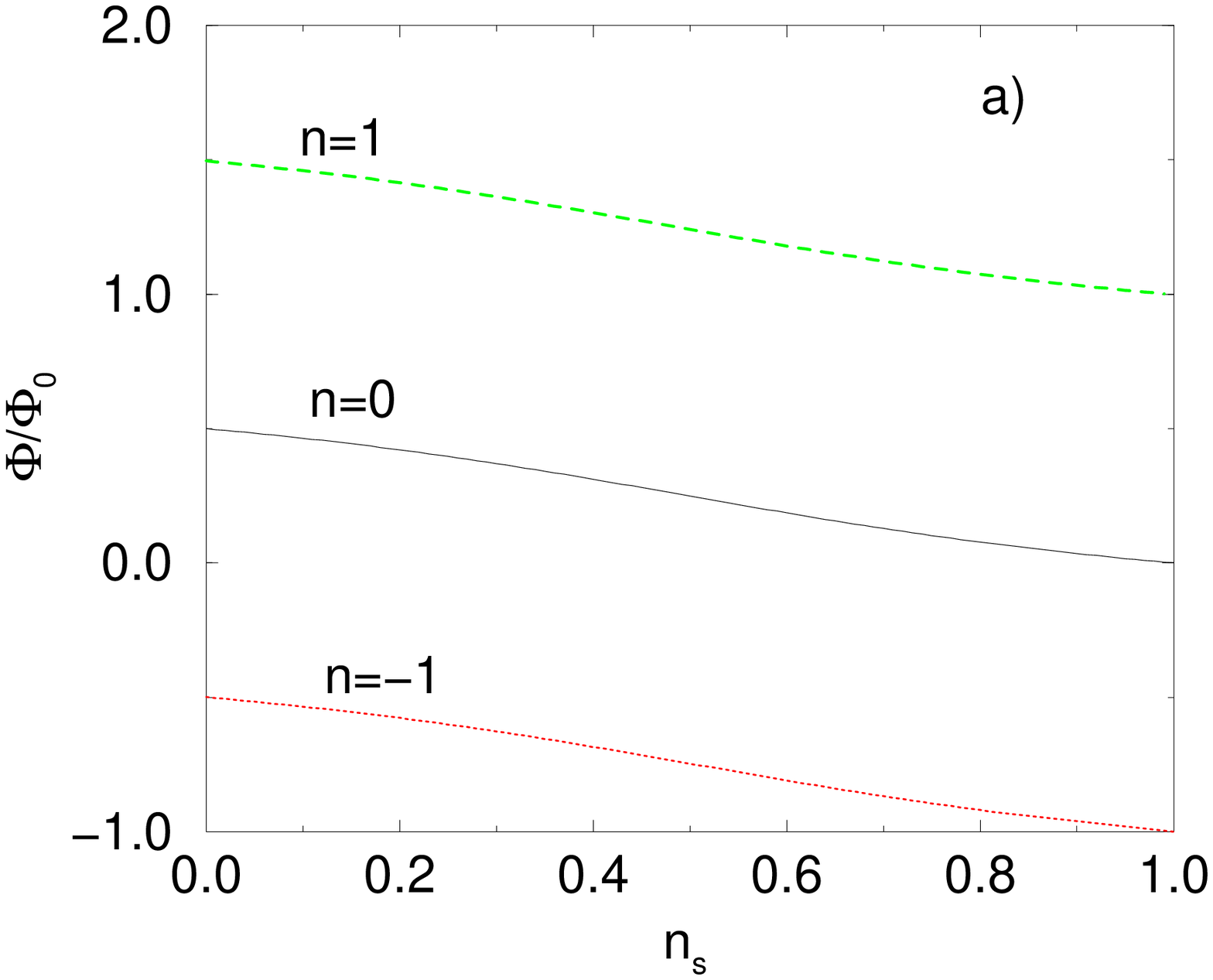,width=6.5cm,angle=0}}
  \centerline{\psfig{figure=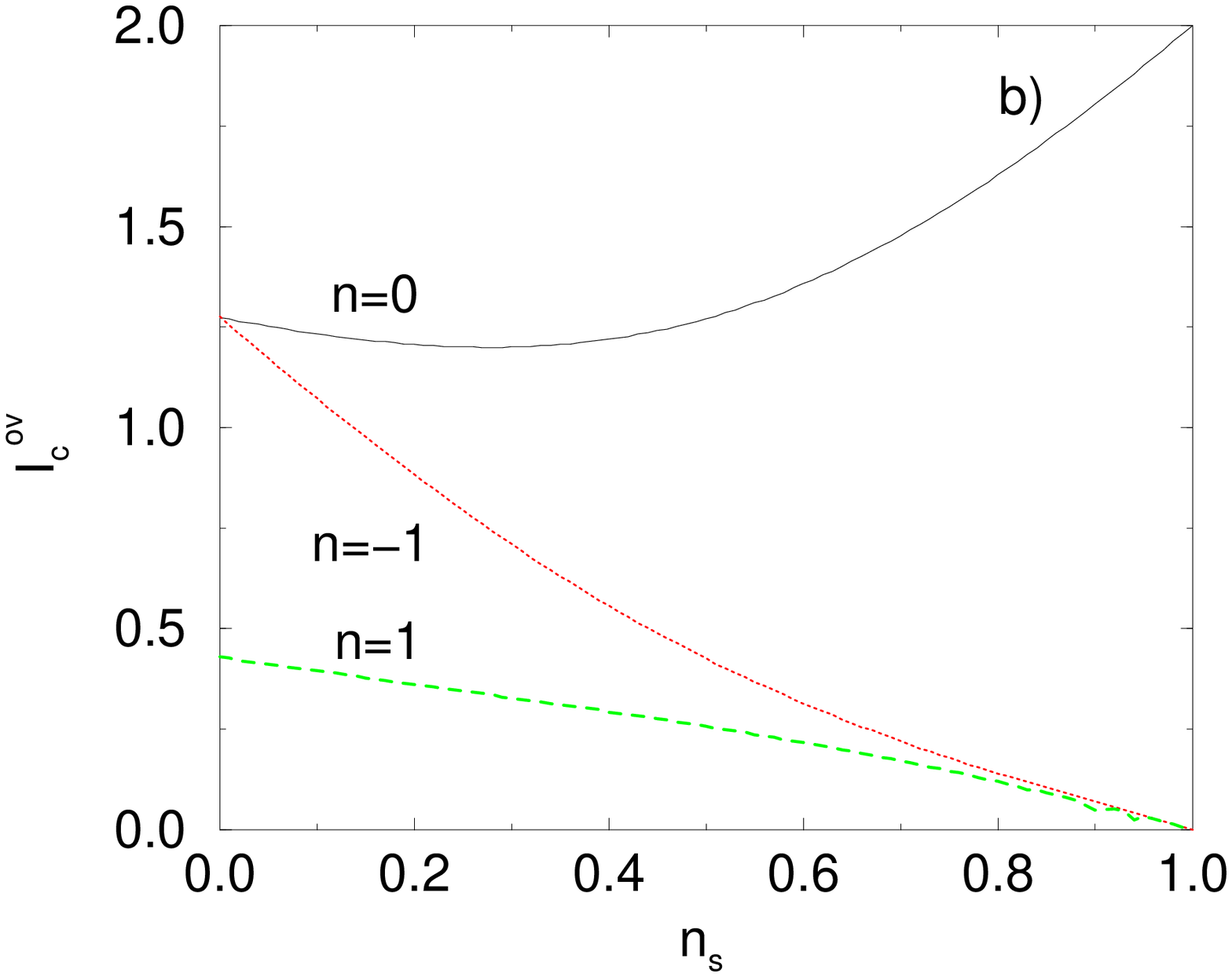,width=6.5cm,angle=0}}
\caption{
a) The spontaneous magnetic flux $\Phi$ 
and b) the critical current $I_c^{ov}$ per unit length 
versus the strength $n_s$ of the 
secondary $s$-wave component
for a corner junction of $d_{x^2-y^2}+is$-wave and 
$s$-wave superconductors,
with length
$L=10\lambda_J$, 
for the vortex solutions $n=0,-1,1$
that exist spontaneously in the junction. The magnitude $n_d$ of the 
$d_{x^2-y^2}$-wave order parameter is given by the relation 
$n_s+n_d=1$.
}
\label{fig9.fig}
\end{figure}

In the above calculations the magnitude of the secondary order
parameter is small compared to the dominant (i.e. $n_{20}=0.1n_{10}$).
However the maximum fraction of the secondary component that has
been observed in phase coherent experiments employing different
materials, geometries, and techniques is up to $25\%$ of the
dominant \cite{vanh}. This triggered our interest to study the magnetic
flux and also the critical currents as a function of the
strength ($n_s$) of the secondary order parameter,
where the magnitude of the dominant order parameter $n_d$ is also
varied in a way that $n_s+n_d=1$. When $n_s=0$ only the $d_{x^2-y^2}$-wave
order parameter is present, while when $n_s=1$ only the $s$-wave
order parameter appears. This situation can be realized
for example near the surface where the $d_{x^2-y^2}$-wave order parameter
is suppressed and the $s$-wave order parameter is enhanced.
The result is presented in Fig. \ref{fig9.fig}a) and \ref{fig9.fig}b)
for the $d_{x^2-y^2}+is$-wave
case at $\theta=0$. We see that when the secondary component
is absent (i.e. $n_s=0$) the picture of the
$d_{x^2-y^2}$-wave state is reproduced. The same picture also
holds for the $d_{x^2-y^2}+id_{xy}$-wave state at $\theta=0$, since
the order parameter for the $d_{x^2-y^2}+id_{xy}$-wave state at
$\theta=0$ is real not breaking the time-reversal symmetry.
So
for $\theta=0$ the magnetic flux and the critical current
for the $d_{x^2-y^2}+id_{xy}$-wave state would not show
any change with the variation of the secondary order parameter
$d_{xy}$.
As $n_s$ is increasing the modes $n=0$ and $n=-1$ are
no more degenerate, in the sense that their flux deviates
from the value $\Phi_0/2$ and $-\Phi_0/2$ respectively
and also their critical currents are no
longer equal. The mode $n=0$ has larger critical current
because it has smaller flux content in absolute value.
For values of $n_s$ close to unity, the different modes
contain integer magnetic flux, as in the junction
between $s$-wave superconductors, and also their
critical currents have the same values
as in the perfect junction problem.
The conclusion is that the larger the secondary component
is in a sample the easier is to be detected in a flux measurement
experiment.

\subsection{Magnetic field}
We now examine the influence of the magnetic field on the spontaneous 
vortices for broken time reversal symmetry pairing states. In Fig. 10 
we plot the magnetic flux at zero current versus the magnetic field $H$
for the $d_{x^2-y^2}+id_{xy}$-wave state at $\theta=0$. In the pure 
$s$-wave superconductor junction
there is no overlap between different modes in the magnetic flux, and 
each mode has magnetic flux which is more than $n\Phi_0$ and less 
than $(n+1)\Phi_0$. In this problem due to spontaneous magnetization 
the range of the modes is different and in some cases overlapping, 
and the labeling
is with a single index $n$, corresponding to the 
pure $s$-wave superconductor 
junction (n,n+1) mode \cite{owen}. Moreover the range in magnetic flux 
of each mode is displaced compared to the pure $s$-wave 
superconductor junction 
problem by an amount which corresponds to the intrinsic flux.
Also we have the existence of 
stable vortex states i.e. $n=0,-1$, together with the unstable ones 
i.e. $n=1$, $-2$ in a large interval of the magnetic field, which 
is almost the same.
The $n=-2$ mode extends to zero magnetic field, and the reason we 
didn't examined this mode in Sec. IV is because the stability analysis
shows negative eigenvalues for all the range of junction orientations, 
at $H=0$.
In the long $s$-wave junction the extremum of the mode $(0,1)$ in $H$ 
is the critical field for one fluxon (antifluxon) penetration from the edges, 
[denoted by $H_{cr}$ $(H_{cl})$, for the right (left) edge], and 
is equal to $2 (-2)$. The solution for the phase at these extremum values 
of the field becomes unstable because the value of the phase at 
the junction edges reaches a critical value. 
In the problem of a junction with some spontaneous flux, we 
consider here, the range of the corresponding mode $0$ in $H$ is 
significantly broadened and also the instability at the boundaries 
sets in due to different reasons. In particular the instability 
occurs due to the interaction of the flux entering from the 
junction edges, when the magnetic field reaches the critical 
value $H_{cr}(H_{cl})$, with the spontaneous flux at the center.
Similar features are encountered in the problem of flux pinning 
from a macroscopic defect in a conventional $s$-wave junction. \cite{defect}

\begin{figure}
  \centerline{\psfig{figure=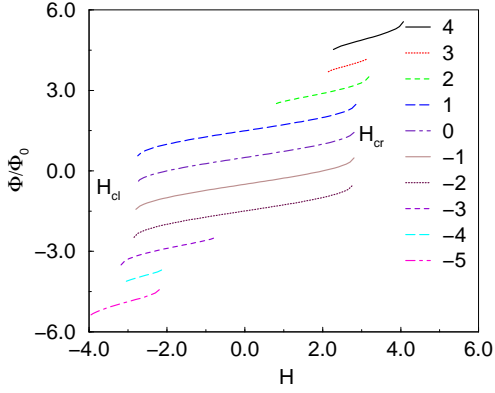,width=6.5cm,angle=0}}
\caption{ Magnetic flux $\Phi / \Phi_0$ at zero external current versus the 
magnetic field $H$ 
for a corner junction of $d_{x^2-y^2}+id_{xy}$-wave and $s$-wave 
superconductors, with length 
$L=10\lambda_J$, for angle $\theta=0^{\circ}$.
$H_{cl}(H_{cr})$ denotes the critical values of the magnetic field 
where the mode $n=0$, terminates. 
}
\label{fig10.fig}
\end{figure}

\begin{figure}
  \centerline{\psfig{figure=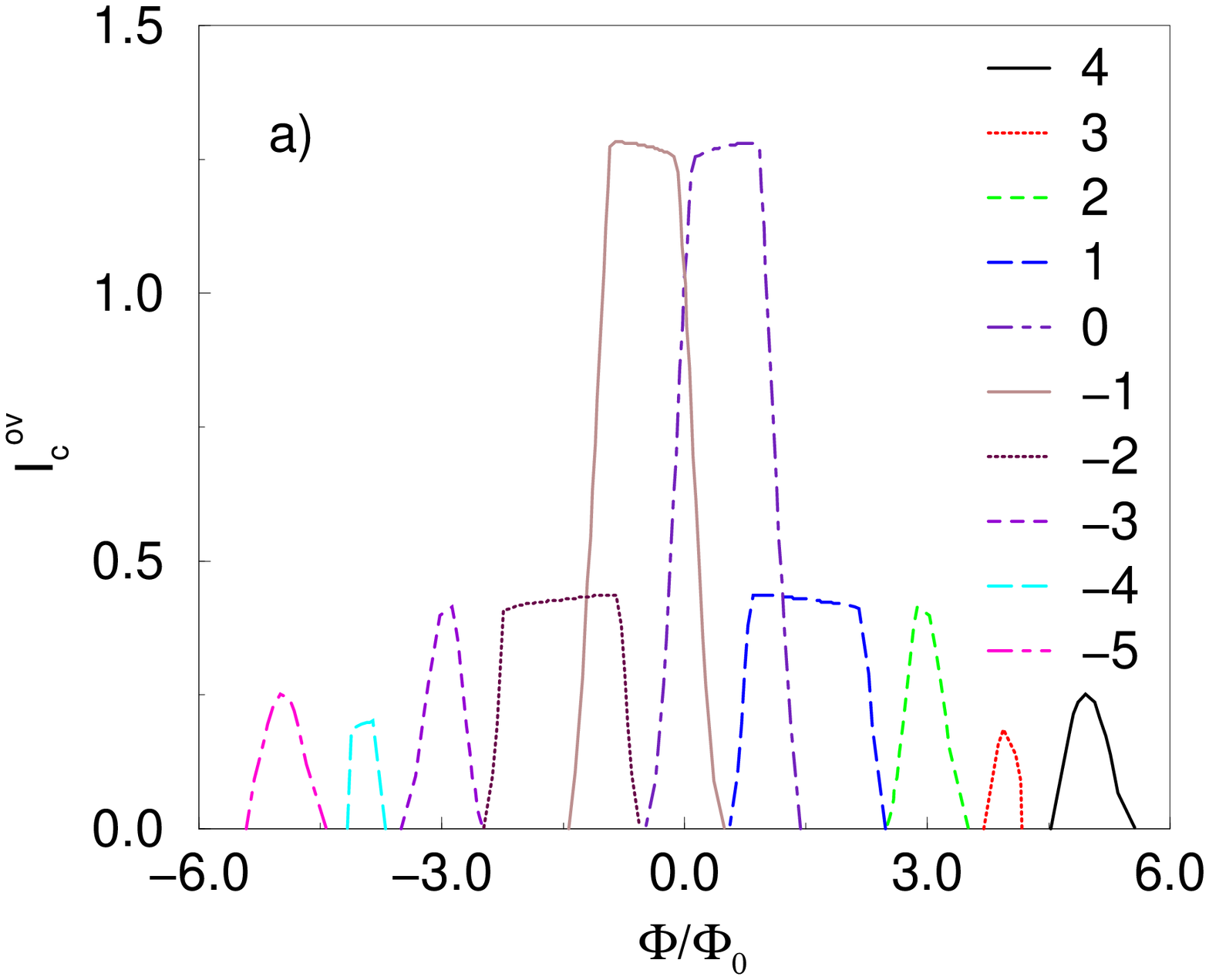,width=6.5cm,angle=0}}
  \centerline{\psfig{figure=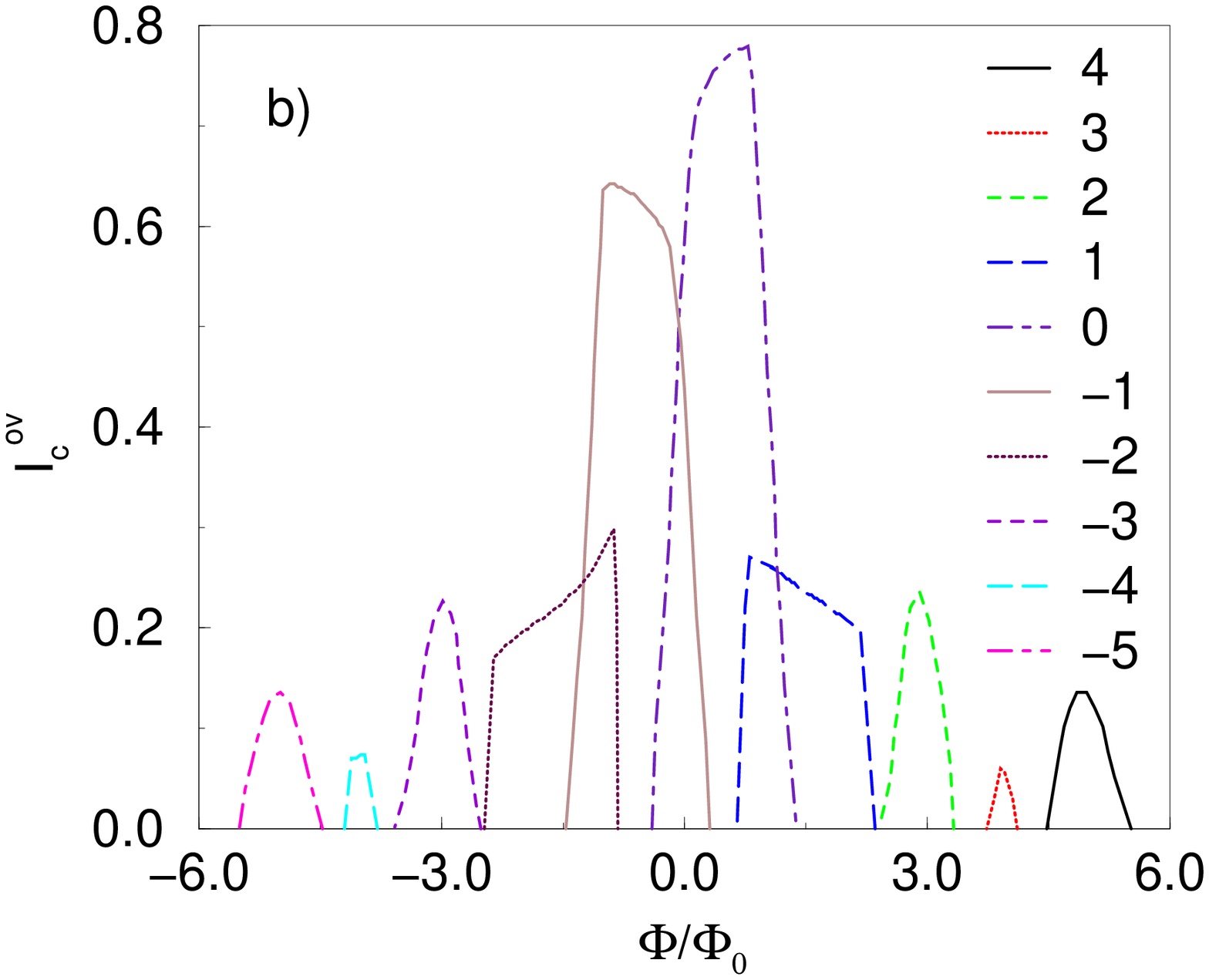,width=6.5cm,angle=0}}
  \centerline{\psfig{figure=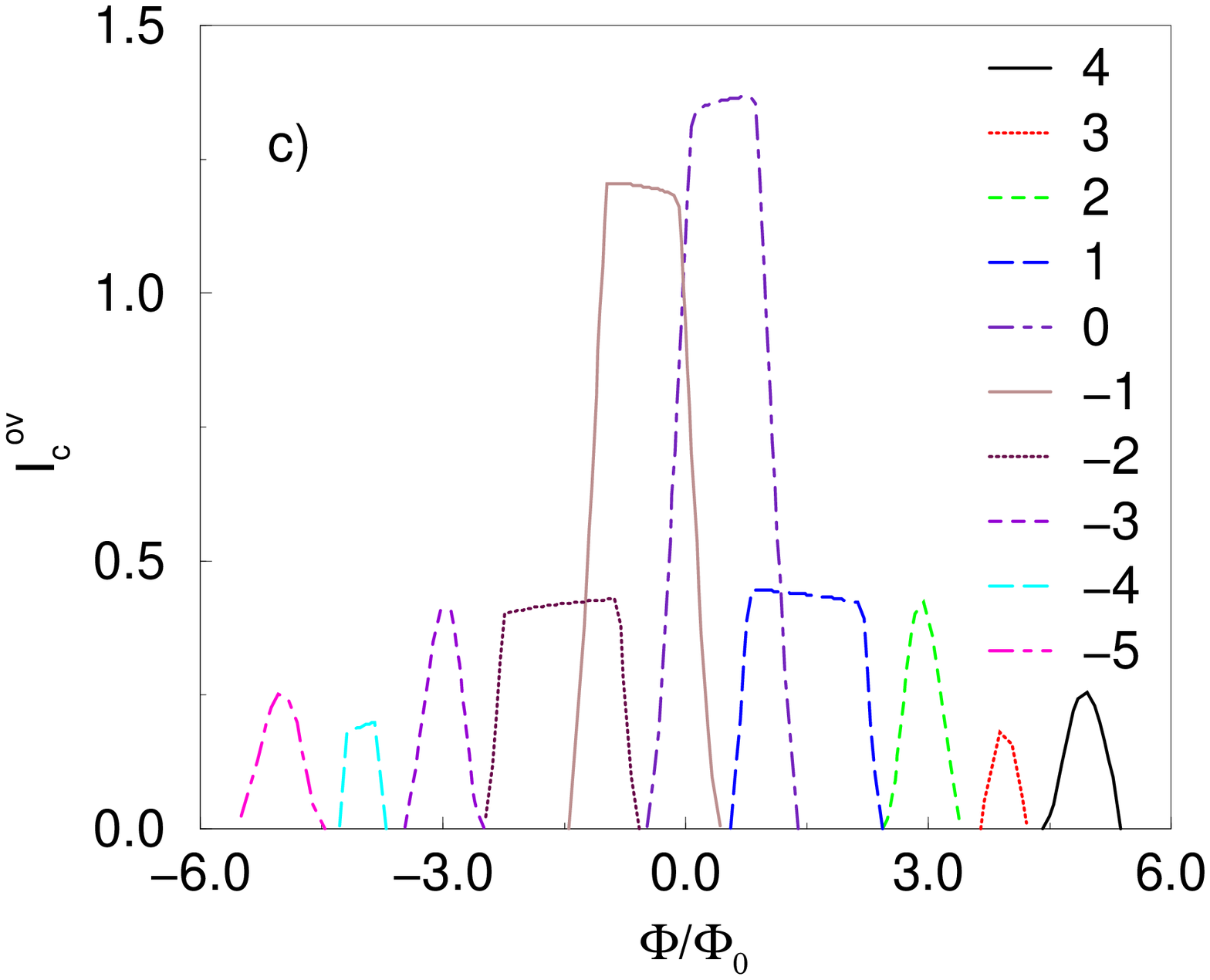,width=6.5cm,angle=0}}
  \centerline{\psfig{figure=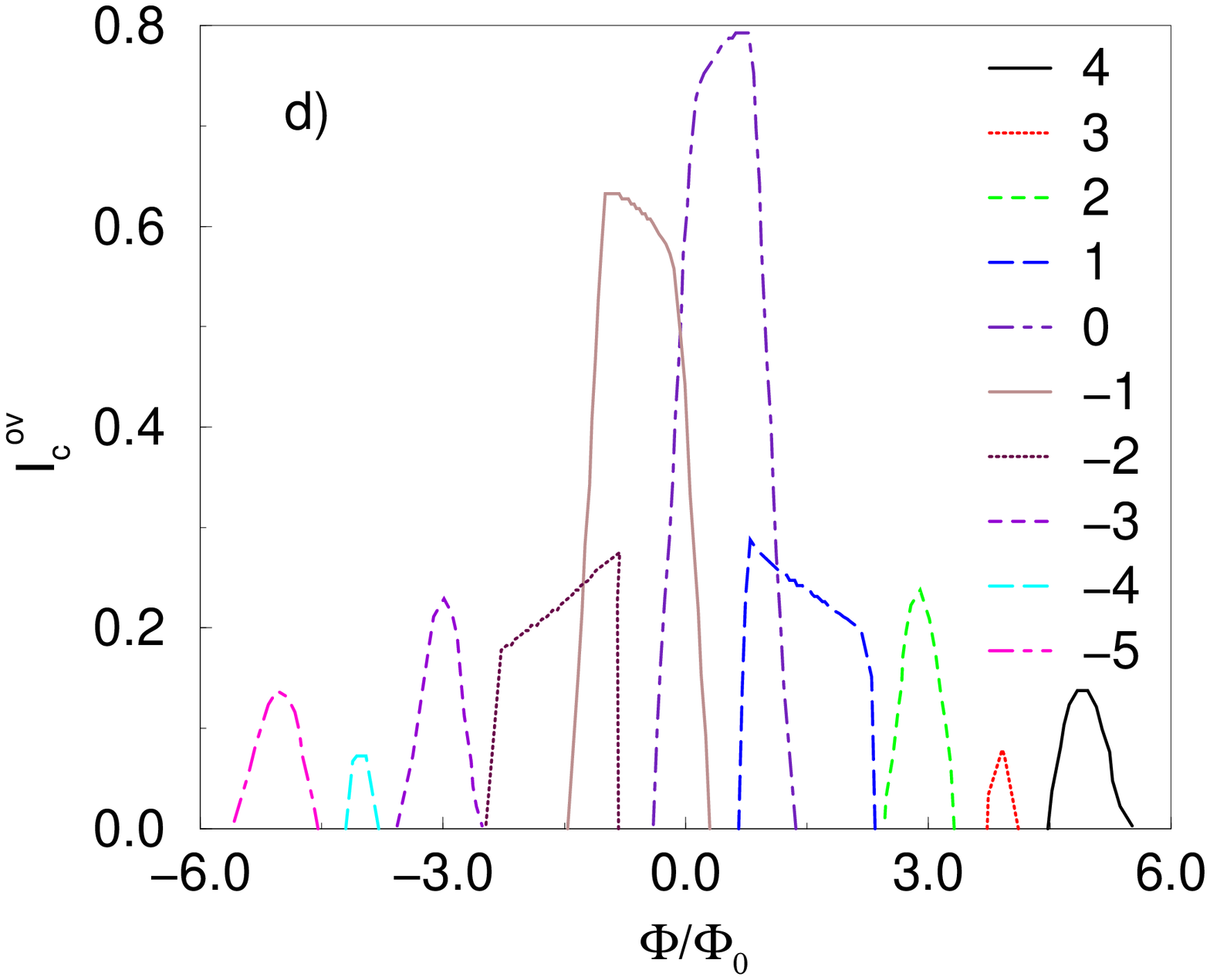,width=6.5cm,angle=0}}
\caption{(a) Overlap critical current $I_c^{ov}$ per unit length versus the 
magnetic flux $\Phi$ in units of $\Phi_0$,
for a corner junction of $d_{x^2-y^2}+id_{xy}$-wave and $s$-wave 
superconductors, with length 
$L=10\lambda_J$, for angle $\theta=0^{\circ}$.
(b) The same as in a) but for $\theta=0.5$.
(c) The same as in a) but for $d_{x^2-y^2}+is$-wave and $s$-wave
superconductors for angle $\theta=0^{\circ}$.
(d) The same as in (c) but for $\theta=0.5$. 
}
\label{fig11.fig}
\end{figure}

We now examine the magnetic-interference pattern for the two symmetries
where the bias current enters in the overlap geometry. 
In the $d_{x^2-y^2}+id_{xy}$-wave case, where $\theta=0$,
this pattern has a symmetric form as we can see 
from Fig. \ref{fig11.fig}(a). 
This is because this result is only sensitive to the real 
part of the order parameter, which has a sign change but 
does not break time-reversal symmetry. 
For the angle $\theta=0.5$ where the order parameter has a 
finite imaginary part and breaks the time-reversal symmetry this 
pattern becomes asymmetric and the ''dip'' appears to a value of 
flux slightly different than zero. Note that the asymmetry refers mainly to the 
modes $n=0$, and $n=-1$. The other modes are not influenced much 
due to their higher flux content. 
Also the critical current is suppressed
compared to the case where $\theta=0$ as can be seen in Fig. 
\ref{fig11.fig}(b), due to a drop in $J_c$. 

In the $d_{x^2-y^2}+is$-wave symmetry, 
in the limit where 
$\theta\rightarrow 0$, the order parameter is complex and 
the pattern is asymmetric 
as can be seen in Fig. \ref{fig11.fig}c, for the
angle $\theta=0$. 
This is in 
agreement with our previous work for the inline current input
for a junction with $d_{x^2-y^2}+is$ symmetry 
\cite{stefan}. 
There it was found that the pattern is asymmetric for lengths as long 
as $L=10\lambda_J$. 
For angles close to $\pi/4$,
the magnetic interference 
pattern is similar with the $d_{x^2-y^2}+id_{xy}$-state.
This is because the $\sin(2\theta)$ dependence of the $d_{xy}$
component is almost unity. 
This is seen in Fig. \ref{fig11.fig}d where we present 
the variation of the critical current per unit length versus the enclosed flux for 
$\theta=0.5$, and the symmetry state is $d_{x^2-y^2}+is$.

In the short junction limit $L<\lambda_J$ the same argument can 
be applied without any explicit reference to fractional
vortex and antivortex solutions. However as we found in our 
previous work \cite{stefan}, both $n=0$ and $n=-1$ (there $f_{va}, f_a$)
exist, with
reduced flux content, in 
this limit as a continuation of the corresponding solutions 
in the large junction limit. In this case the external applied 
magnetic field becomes equal to the self field, and the maximum 
current can be calculated analytically \cite{zhu}, 
\begin{equation}
\frac{I_m(\Phi)}{I_{m0}}=\left|\frac{\sin(\pi \Phi /2\Phi_0) \cos[\pi \Phi/2\Phi_0 + 
(\phi_{c2}-\phi_{c1})/2]}{\pi\Phi/2\Phi_0}\right|
.~~~\label{short}
\end{equation}
As we see at $\theta=0$ for the $d_{x^2-y^2}+id_{xy}$-wave case, the relation 
$\phi_{c2}-\phi_{c1}=n\pi$ holds, and the magnetic interference 
pattern becomes symmetric, while for the $d_{x^2-y^2}+is$, 
this difference is a fraction of $\pi$ and the pattern is asymmetric.
However as we increase the junction length, we expect this symmetric 
pattern for the $d$-wave order parameter to be continued. 
This symmetry in the large junction limit, is described more effectively 
by the assumption of the $n=0,-1$ solutions which give a 
symmetric magnetic interference pattern as we presented. 
Also the $n=-1$ solution extends to values for the 
magnetic flux, where the 
$n=0$ solution is absent. Eliminating one of them will break the 
symmetry of the diagram.

\section{Experimental relevance} 
The symmetric pattern with a minimum at zero applied field 
observed in corner junction experiments between YBCO and Pb 
at $\theta=0$ has been interpreted as an indication of 
$d_{x^2-y^2}$-wave symmetry. \cite{wollman,wollman1}
This result refers to short junctions 
where the junction size is much smaller than the Josephson penetration 
depth. However as we found here these experimental data are also
consistent with an order parameter with $d_{x^2-y^2}+id_{xy}$
pairing symmetry at $\theta=0$. 

Also the critical current $I_c$ 
versus the magnetic flux $\Phi$ of a SQUID, consisting of two planar
Josephson junctions on the faces of YBCO superconducting crystal, 
connected by a loop of a second superconductor, 
for $\theta=0$ or $\theta=\pi /2$
is found shifted by
$\Phi=0.5\Phi_0$ and has a minimum at $\Phi=0$ 
(instead of a maximum as in a SQUID involving conventional 
$s$-wave superconductors or the edge SQUID in which both junctions are 
on the same crystal face) but is still symmetric.
This result has been attributed to an order parameter 
with $d_{x^2-y^2}$-wave symmetry. However the 
theoretical analysis done by Beasley et al. \cite{beasley} 
shows that it is also consistent with an order parameter 
with $d_{x^2-y^2}+id_{xy}$-pairing symmetry at $\theta=0$. 

In both cases of SQUID and corner junction the symmetric pattern
observed at $\theta=0$ rules out the $d_{x^2-y^2}+is$-wave 
pairing state where the order parameter is complex everywhere 
resulting in an asymmetric $I_c$ versus $\Phi$ pattern for all 
angles $\theta$. 
However the small asymmetry (less than $2\%$) observed at $\theta=0$
in some experiments can be attributed to various complicating 
factors e.g. fluxon trapping as will be discussed latter in  
this section.

The experiment proposed here to
resolve ambiguity between $d_{x^2-y^2}+id_{xy}$ and 
$d_{x^2-y^2}$ at $\theta=0$, is to execute the 
same experiments using SQUID or corner junction 
at an angle between sample faces $\theta$ between $0$ and $\pi /2$. 
Our theory predicts symmetric(asymmetric) pattern for the 
$d_{x^2-y^2}$-wave ($d_{x^2-y^2}+id_{xy}$)-wave pairing state for 
the corner junction case.
This kind of experiment has already been done in the case of 
SQUID geometry. \cite{yanoff}
The tunneling directions are defined lithographically and 
patterned by ion milling of a $c$-axis oriented film. 
A YBCO thin film is patterned into a circle with a series of 
Nb-Au-YBCO edge junctions at orientations spaces every $7.5^\circ$. 
The measurement of the $I_c$ vs $\theta$, which probes mainly the 
magnitude of the order parameter has an angular anisotropy, indicating 
an anisotropic order parameter. Also the execution of this 
experiments is not easy due to the difficulty in cleaning, polishing 
a crystal at angle $\theta$, between $0$ and $\pi /2$.

Also in an experiment analogous to the corner junction 
Miller, Ying et. al. \cite{miller} used frustrated thin-film tricrystal 
samples to probe the pairing symmetry of YBCO. They found a minimum 
in the $I_c$ vs the  externally applied flux 
$\Phi_e$ diagram at $\Phi_e=0$ in the short junction limit
and a maximum at $\Phi_e=0$ for a wide junction where the junction 
length is much larger than the $\lambda_J$. However 
for a wide junction the correct quantity to be compared should 
be the total flux $\Phi$ which involves contribution both 
from the externally applied flux and the intrinsic flux. 
Also in the tricrystal magnetometry experiments on half-flux 
quantum Josephson vorticies one can only observe spontaneous 
magnetization of $\Phi_0/2$ in a frustrated geometry only in the 
large junction length limit \cite{kirtley}.
 
There is a number of complicating factors in the interpretation of the 
experiments involving corner junctions that could lead to 
an asymmetric ($I_c$ vs $\Phi$) pattern even for $\theta=0$. These are 
the asymmetry of the junction (meaning that the 
critical current of the two junction faces are not equal). 
This will only cause the dip 
to be shallower and will maintain the symmetry of the $I_c$ 
vs $\Phi$ diagram. 
Also these experiments are influenced by the sample geometry and 
the effect of flux 
trapping i.e. there can be vortices trapped between the planes
of the cuprate superconductors that could affect the $I_c$ vs $\Phi$ diagram.
In the corner junction case, it creates an asymmetry in the 
flux modulation curves.
However these flux trapping effects are not sufficiently large 
to change the qualitative interpretation of these experiments.

\section{Conclusions} 
We studied numerically the possible spontaneous vortex states
that may exist in a corner junction between a superconductor with 
time reversal symmetry broken, (i.e. $d_{x^2-y^2}+id_{xy}$ 
or $d_{x^2-y^2}+is$), and an $s$-wave superconductor, 
in 
the long junction limit. 
We studied separately three parameters which can be used 
to modulate the spontaneous flux. These are the magnetic field 
$H$, the interface orientation $\theta$, and the magnitude of 
the subdominant order parameter $n_s$. We pointed out the differences 
between time reversal broken states under these modulation 
parameters.

We found that in flux modulation experiments involving superconductors 
with some spontaneous flux the range in
magnetic flux of each mode is displaced compared to the 
case of a pure $s$-wave superconductor junction by an amount which corresponds 
to the intrinsic flux. 
In particular when the magnetic field $H$ is considered as the 
modulation parameter, the range in $H$ of the lower fluxon modes 
is significantly broadened compared to the $s$-wave case, and 
the instability at the boundary values of the field sets in 
due to the interaction of the flux entering from the 
junction edges with the intrinsic flux.
In any case, for each value of the parameter 
which changes the flux,  
the modes are 
separated by a single flux quantum.

We also derived some simple arguments to discriminate between 
the different pairing states that break the time reversal symmetry.
For the $d_{x^2-y^2}+id_{xy}$-wave pairing state, the 
junction orientation where $\theta=0$ i.e. the lobes of the 
dominant $d_{x^2-y^2}$-wave order parameter are at right angles 
for the corner junction, give flux quantization condition 
$\Phi=n\Phi_0/2$ as in 
the $d_{x^2-y^2}$-wave state, which is different from the 
corresponding flux quantization for the $d_{x^2-y^2}+is$-wave 
pairing state, at $\theta=0$, which is $\Phi=(n/2+f)\Phi_0$, 
where $f$ is a small quantity. 
These different conditions provide a way experimentally to 
distinguish between time reversal broken symmetry states.
Note that since the magnitude of the secondary order 
parameter is small compared to the dominant, the detection of 
time reversal broken states requires a very precise measurement 
of the spontaneous magnetic flux.

Also we showed that the magnetic interference pattern at $\theta=0$ 
is symmetric
(asymmetric) for the $d_{x^2-y^2}+id_{xy}$ ($d_{x^2-y^2}+is$), and 
this also can be used to probe which symmetry the order 
parameter has, at least where the junctions are formed.
We expect our findings, for the magnetic field dependence of the 
critical current, to hold even in the short junction limit, where 
the most experiments on corner junctions have been performed 
\cite{vanh,yanoff}. 

\section{Acknowledgments}
One of us N.S. is grateful to A.V. Balatsky, J. Betouras for useful discussions 
that led to this article. Also N.S. would like to acknowledge 
the ESF/FERLIN programme for partial support to participate 
to conferences.


\begin{references}
\bibitem{scalapino} D.J. Scalapino, Phys. Rep. {\bf 250}, 329 (1995).

\bibitem{vanh} D.J. van Harlingen, Rev. Mod. Phys. {\bf 67}, 515 (1995).

\bibitem{shiba} M. Matsumoto, and H. Shiba, J. Phys. Soc. Jpn.
{\bf 64}, 1703 (1995).

\bibitem{kirtley2} J.R. Kirtley, P. Chaudhari, M.B. Ketchen, N. Khare, 
S.Y. Lin, and T. Shaw, Phys. Rev. B {\bf 51}, 12 057 (1995).

\bibitem{bailey} D.B. Bailey, M. Sigrist, and R.B. Laughlin, Phys. Rev. 
B {\bf 55}, 15 239 (1997). 

\bibitem{balatsky1} A.V. Balatsky, Phys. Rev. Lett. {\bf 80}, 1972 (1998).

\bibitem{balatsky2} A.V. Balatsky, Phys. Rev. B {\bf 61}, 6940 (2000).

\bibitem{covington} M. Covington, M. Aprili, E. Paraoanu, L.H. Greene, 
F. Xu, J. Zhu, and C.A. Mirkin, Phys. Rev. Lett. {\bf 79}, 277 (1997).

\bibitem{fogelstrom} M. Fogelstrom, D. Rainer, and J.A. Sauls, 
Phys. Rev. Lett. {\bf 79}, 281 (1997).

\bibitem{aprili} M. Aprili, E. Badica, and L.H. Greene, 
Phys. Rev. Lett. {\bf 83}, 4630 (1999).

\bibitem{krupke} R. Krupke, and G. Deutscher,
Phys. Rev. Lett. {\bf 83}, 4634 (1999).

\bibitem{yanoff} D.J. van Harlingen, J.E. Hilliard, B.L.T. Plourde and 
B.D. Yanoff, Physica C {\bf 317-318}, 410 (1999).

\bibitem{zhu} J.-X. Zhu, W. Kim, and C.S. Ting, 
Phys. Rev. B {\bf 58}, 6455 (1998).

\bibitem{sigrist} M. Sigrist, Prog. Theor. Phys.
{\bf 99}, 899 (1998).

\bibitem{caputo} 
J.-G. Caputo, N. Flytzanis, Y. Gaididei, N. Stefanakis, and 
E. Vavalis, Supercond. Sci. Technol. {\bf 13}, 423 (2000).

\bibitem{owen} C.S. Owen, and D.J. Scalapino, 
Phys. Rev. {\bf 164}, 538 (1967).

\bibitem{defect} N. Stefanakis, and N. Flytzanis, 
Supercond. Sci. Technol. {\bf 14}, 16 (2001).

\bibitem{stefan} N. Stefanakis, and N. Flytzanis, Phys. Rev. 
B {\bf 61}, 4270 (2000). 

\bibitem{wollman} D.A. Wollman, D.J. van Harlingen, W.C. Lee, 
D.M. Ginsberg, and A.J. Leggett, Phys. Rev. Lett. 
{\bf 71}, 2134 (1993). 

\bibitem{wollman1} D.A. Wollman, D.J. van Harlingen, J. Giapintzakis, 
and D.M. Ginsberg, Phys. Rev. Lett. 
{\bf 74}, 797 (1995). 

\bibitem{beasley} M.R. Beasley, D. Lew, and R.B. Laughlin, Phys. Rev. 
B {\bf 49}, 12 330 (1994). 

\bibitem{miller} J.H. Miller Jr., Q.Y. Ying, Z.G. Zou, N.Q. Fan, 
J.H. Xu, M.F. Davis, and J.C. Wolfe, 
Phys. Rev. Lett. 
{\bf 74}, 2347 (1995). 

\bibitem{kirtley} J.R. Kirtley, C.C. Tsuei, J.Z. Sun, L.S. Yu-Jahnes, 
A. Gupta, M.B. Ketchen, K.A. Moler, and M. Bhushan,
Phys. Rev. Lett. {\bf 76}, 1336 (1996).

\end{references}
\end{document}